\documentclass[fleqn,usenatbib]{mnras}

\usepackage{newtxtext,newtxmath}

\usepackage[T1]{fontenc}

\DeclareRobustCommand{\VAN}[3]{#2}
\let\VANthebibliography\thebibliography
\def\thebibliography{\DeclareRobustCommand{\VAN}[3]{##3}\VANthebibliography}

\usepackage{graphicx}	
\usepackage{amsmath}	
\usepackage{amssymb}	

\title[No evidence of MPs in MS of NGC 419]{Searching for globular cluster chemical anomalies on the main sequence of a young massive cluster}

\author[Cabrera-Ziri et al.]{
I. Cabrera-Ziri,$^{1}$\thanks{Hubble Fellow}\thanks{E-mail: ivan.cabrera@cfa.harvard.edu}
J. S. Speagle,$^{1}$  
E. Dalessandro,$^{2}$  
C. Usher,$^{3,4}$ 
N. J. Bastian,$^{3}$ \linebreak
\newauthor
M. Salaris,$^{3}$ 
S. Martocchia,$^{3,5}$
V. Kozhurina-Platais,$^{6}$
F. Niederhofer,$^{7}$
C. Lardo,$^{8}$\linebreak
\newauthor
S. S. Larsen,$^{9}$
S. Saracino$^{3}$\\
$^{1}$Center for Astrophysics | Harvard \& Smithsonian, 60 Garden Street, Cambridge, MA 02138, USA\\
$^{2}$INAF-Osservatorio di Astrofisica e Scienza dello Spazio di Bologna, Via Gobetti 93/3, Bologna, II-40129, Italy\\
$^{3}$Astrophysics Research Institute, Liverpool John Moores University, 146 Brownlow Hill, Liverpool L3 5RF, UK\\
$^{4}$Department of Astronomy, Oskar Klein Centre, Stockholm University, AlbaNova University Centre, SE-106 91 Stockholm, Sweden\\
$^{5}$European Southern Observatory, Karl-Schwarzschild-Strasse 2, D-85748 Garching bei Munchen, Germany\\
$^{6}$Space Telescope Science Institute, 3700 San Martin Drive, Baltimore, MD 21218, USA\\
$^{7}$Leibniz-Institut f\"ur Astrophysik Potsdam (AIP), An der Sternwarte 16, D-14482 Potsdam, Germany\\
$^{8}$Laboratoire d'Astrophysique, Ecole Polytechnique F\'ed\'erale de Lausanne, Observatoire de Sauverny, CH-1290 Versoix, CH\\
$^{9}$Department of Astrophysics/IMAPP, Radboud University, PO Box 9010, 6500 GL Nijmegen, The Netherlands
}

\date{Accepted 2020 April 20. Received 2020 April 19; in original form 2020 February 7}

\pubyear{2020}

\begin{document}
\label{firstpage}
\pagerange{\pageref{firstpage}--\pageref{lastpage}}
\maketitle

\begin{abstract}
The spectroscopic and photometric signals of the star-to-star abundance variations found in globular clusters seem to be correlated with global parameters like the cluster's metallicity, mass and age. Understanding this behaviour could bring us closer to the origin of these intriguing abundance spreads. In this work we use deep \emph{HST} photometry to look for evidence of abundance variations in the main sequence of a young massive cluster NGC 419 ($\sim10^5$ M$_{\odot}$, $\sim1.4$ Gyr). Unlike previous studies, \emph{here we focus on stars in the same mass range found in old globulars} ($\sim0.75-1$ M$_{\odot}$), \emph{where light elements variations are detected}. We find no evidence for N abundance variations among these stars in the $Un-B$ and $U-B$ CMD of NGC 419. This is at odds with the N-variations found in old globulars like 47 Tuc, NGC 6352 and NGC 6637 with similar metallicity to NGC 419. Although the signature of the abundance variations characteristic of old globulars appears to be significantly smaller or absent in this young cluster, we cannot conclude if this effect is mainly driven by its age or its mass.
\end{abstract}




\section{Introduction}
\label{sec:intro}

The puzzle that is the origin of the star-to-star light element abundance variations within globular clusters (a.k.a. multiple stellar populations, MPs) is still unsolved. Since the beginning, an interesting piece of this puzzle has been the fact that clusters with similar metallicity can display very different ranges of their light element abundances variations \cite[e.g.][]{Kraft79}. 
Large spectroscopic campaigns in the last decade have brought us a better understanding of the behaviour of some light elements as a function of global parameters, e.g. \cite{Carretta10} found a correlation between the absolute magnitude of globular clusters (a proxy of their current mass) and the extent of some abundance distributions.

Similar findings have been produced by photometric studies using special filter combinations (a.k.a. pseudocolours or supercolours) involving near-UV and/or blue ($<4000$~\AA) filters which are able to pick up variations in the chemistry of the atmospheres (namely C, N and O) and structure of the stars (consequence of different He mass fractions), e.g. \cite{Monelli13}. Arguably, \cite{Milone17} with the use of \emph{HST} supercolours, provided the best picture to date of the manifestation of MPs in Galactic globulars as a function of global cluster properties like mass and metallicity. Understanding the physics of how such global parameters can regulate the expression of MPs at individual star level, e.g. how severe are the relative difference in the abundances of stars within a given cluster, could provide valuable insights into the origin of this phenomenon.

The role that age plays in the way MP manifest has been difficult to constrain with Galactic targets. Studies of young open clusters have not found evidence for significant abundance variations among their stars (e.g. \citealt{deSilva09,Smiljanic09,Pancino10a,Carrera11,Carrera13,MacLean15}). However, attributing this to an age effect is not straight forward for a couple of reasons: 1) there is a strong correlation between the cluster mass and the degree of change in light element abundance among cluster stars \cite[e.g.][]{Schiavon13,Milone17} and 2) metallicity is also known to modulate the signatures of MPs, with results suggesting that (on average) the abundance spread in metal rich clusters is smaller than metal poor ones \cite[e.g.][]{Pancino17,Meszaros20}.

These open clusters are several orders of magnitude less massive than the typical old globular and, their metallicity is on average significantly higher (i.e. around solar instead of [Fe/H] $\lesssim -0.7$ dex). Both factors seem to attenuate the signal of the MPs found in Galactic globulars hindering any conclusions about the role age might play in the MP phenomenon. 
Fortunately, massive clusters have been forming in the Magellanic Clouds almost continuously for a Hubble time \cite[cf.][]{Glatt09,Baumgardt13}. As these clusters are relatively metal poor, they provide the best opportunity to explore how the MPs express at different ages in a more controlled way.

Using a sample of Magellanic Clouds clusters \cite{Martocchia18} showed that at a given cluster mass ($\sim 10^5$ M$_{\odot}$) the width of the RGB in CMDs when using supercolours increases as a function of cluster age, suggesting the abundance variations present in the RGBs of older clusters are more severe than the ones found in the younger ones. Similar  results are presented by \cite{Lagioia19} in their Fig. 10. These photometric findings are in agreement with spectroscopic studies like \cite{Hollyhead19} showing N-variations $<1$~dex at young ages ($\sim2$~Gyr) compared to the ones characteristic of old Galactic globulars which can comfortably exceed one dex (e.g. \citealt{Cohen05,Martell09,Pancino10b}).

However, \cite{Salaris20} have shown that for RGB stars the comparison of the signals of N-variations is not straight forward between clusters of different ages due to changes in the efficiency of the first dredge-up. Essentially, for a constant initial star-to-star N-abundance difference, the  observed N difference between sub-populations would appear to increase for increasing age.  While this qualitatively matches observations, \citeauthor{Salaris20} found that the observed increase is stronger than the effect of the first dredge-up in their models, implying an intrinsic increase in N-abundance with cluster age. 
However it is clear that, evolutionary effects like this, hamper any conclusions coming out of RGB stars regarding the role the cluster age plays in the \emph{initial} star-to-star abundance variation.

To investigate the effect of this issue, in this paper we perform the critical step forward to study light-element abundance variations on the main sequence (MS), whose stars are not affected by evolutionary effects and preserve their initial chemical composition, and also avoiding the effects of stellar rotation.
These abundance variations have been traced down to $\sim0.2$~M$_{\odot}$~in old globulars, e.g. \cite{Milone12}. In this work we use deep imaging of NGC 419, a massive ($\sim 10^5$~M$_{\odot}$, \citealt{Kamann18}), young ($\sim1.4$~Gyr, \citealt{Glatt09}) Small Magellanic Cloud cluster with $\mbox{[Fe/H]}\sim-0.7$~dex, in order to \emph{search for MPs in stars in the same mass range where they are found in old globulars}
 (i.e. $\lesssim1$~M$_{\odot}$~stars). The presence or absence of MPs in these stars will produce a clearer picture regarding the role cluster age plays in the manifestation of this phenomenon.

\section{Data and models}
\label{sec:dat}

For NGC 419, we used $HST$ images taken with the ACS/WFC and WFC3/UVIS cameras in bands $F336W$, $F343N$, $F438W$, $F555W$ and $F814W$\footnote{For simplicity we will refer to them as: $U$, $Un$, $B$, $V$ and $I$ respectively.} for our CMD analysis (Programmes GO-10396, GO-12257, GO-14069 and GO-15061). We performed PSF photometry and artificial star tests (AST) using \texttt{DAOPHOT II} and \texttt{ALLFRAME} \citep{Stetson87,Stetson94} following the same procedure described in \cite{Bellazzini02,Dalessandro15,Martocchia18} and references therein. For our study we focused on the stars within the half-light radius of NGC 419. We decontaminated the CMDs following the procedure outlined in Appendix \ref{sec:decon} (cf. Fig. \ref{fig:cleaning_1ite}).

We use as reference clusters 47 Tuc (NGC 104), NGC 6352 and NGC 6637, all of them old Galactic globulars with a similar metallicity to NGC 419 (more on this in \S\ref{sec:ms}). The photometry for these clusters was taken from (cf. \citealt{Nardiello18}, N18 from now on). The catalogues were cleaned by setting a $p>95\%$ cut in their membership probability based on their proper motions (cf. N18). Finally, we supplemented the N18 photometry of 47 Tuc with $Un$ band and the deeper $B$ from HST programme GO-15061. The $Un$ band was not available for the other clusters in our sample.

We also make use the isochrones presented in \citet{Martocchia17} on their analysis of NGC 419. These are 1.41 Gyr, [Fe/H] $= -0.7$ and include different chemical composition, namely: a scaled solar model ([C/Fe] = [N/Fe] = [O/Fe] $= 0.0$); an intermediately N-enhanced model ([C/Fe] = [O/Fe] $= -0.1$, [N/Fe] $= +0.5$); and an `extremely' N-enhanced model ([C/Fe] = [O/Fe] $= -0.6$, [N/Fe] $= +1.0$).

\section{Analysis}
\label{sec:ms}

As mentioned above, our goal is to search for abundance variations characteristic of older clusters (i.e. MPs) using colours that are mostly sensitive to changes in N. Although stars in the main sequence have not experienced the first-dredge-up, we still need to proceed with care when selecting the sample of stars for this analysis to make sure that the signal of the N-variations is not affected by different phenomena (at least in a significant way).

\begin{figure}
\begin{center}    
	\includegraphics[height=7.5cm]{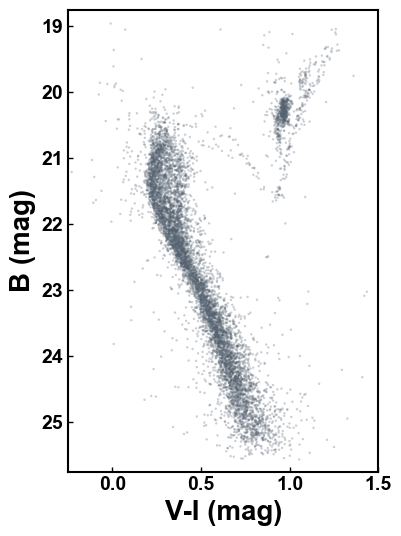}
    \caption{V-I vs. B CMD of NGC 419. The presence of fast rotating stars in this cluster can be inferred from the shape of the turn-off, see text. The N-surface abundance of these stars might not reflect the initial value (see text). We will avoid these stars for our analysis.}
    \label{fig:cmd}

\end{center}
\end{figure}

The CMD of NGC 419 reveals a great example of an extended MS turn-off (cf. Fig. \ref{fig:cmd}). This feature is a prediction of stellar evolutionary models that include fast rotation rates  \citep[e.g.][]{Brandt15,Niederhofer15,Georgy19,Gossage19}\footnote{Alternative (and less successful) interpretations of the extended MS turn-off phenomenon included: age spreads, abundance variations and variable stars. Although each of these hypothesis had their merits, they all lacked the predictive power of the (confirmed) fast-stellar rotation hypothesis. See \cite{cz18} for a brief review.}. Rotationally induced mixing, brings more Hydrogen to the core of the stars (thus extending their MS lifetime), while the changes in the structure of the stars produce temperature gradients from the equator to the poles. The combination of both effects (prolonged MS time and a temperature dependence of the viewing angle of a star) introduce scatter in magnitude and colour in the turn-off of populations of stars with different rotation rates.
The presence of large fractions of fast rotators in such clusters has been confirmed by a wide collection of studies \citep[e.g.][the latter targeting NGC 419]{Bastian17,Dupree17,Milone18,Kamann18}.

Massive ($\gtrsim1.3$~M$_{\odot}$~stars) main sequence stars that have been rotating fast will have altered surface abundances consequence of rotationally induced mixing (e.g. \citealt{Hunter08,Lagarde12}). This would affect elements like N (qualitatively) and cause N abundance variations similar to those seen in globular clusters
  -- which do not seem to be explained by fast rotation \citep[cf.][]{cz18,BL18}.

\begin{figure}
	\includegraphics[width=\columnwidth]{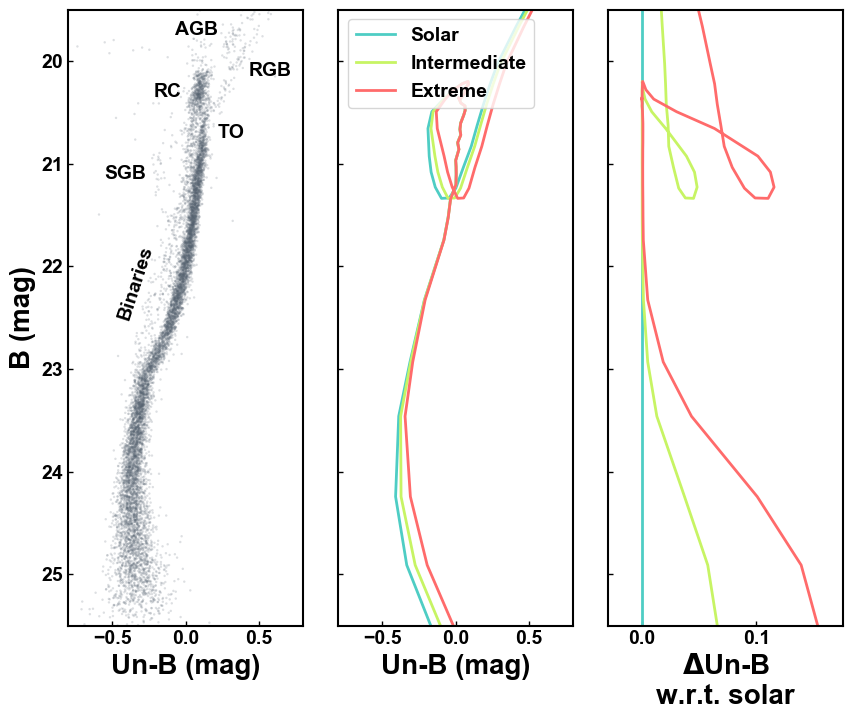}
    \caption{\textbf{Left}: A zoom into the MS of the $Un-B$ vs. $B$ CMD of NGC 419. For clarity we have identified some the main evolutionary stages. \textbf{Middle}: Behaviour of models with different N enrichment in the $Un-B$ CMD. The blue, green and red lines show the predictions of our isochrones with solar scaled, intermediate and extreme N enhancement composition, respectively (cf. \S \ref{sec:dat}). \textbf{Right}: Colour difference between the solar scale and N-enhanced isochrones as a function of magnitude. For $B>22$~mag, the models stars to diverge in the $U-B$ and $Un-B$ colours}.
    \label{fig:model_example}
\end{figure}

Hence we will focus our analysis on stars that should not have been affected by rotationally induced mixing (i.e. magnetically braked stars\footnote{Stars with convective envelopes develop magnetic fields. These cause the wind to rotate as a solid body, transporting angular momentum outwards and slowing the rotation rate, cf. \citealt{Kraft67}.}) and that lie in parts of the CMD where stars of different chemical composition diverge significantly when using the right combination of filters. In Fig.~\ref{fig:model_example} we can see that the isochrones with different composition start separating at magnitudes fainter than $B\sim22$~mag. For our study, we will focus on the stars in the $B$-magnitude range from $23.5-25.5$ mag, which correspond to the mass range between $1.05\gtrsim \mbox{M}_{\odot} \gtrsim 0.75$, according to our models.

We should note that we only use these models to inform us when populations with difference N-variations start to become distinct. These models are not adequate for a direct comparison to the data in order to infer the presence/absence of MPs. For that we will use clusters with known MPs as empirical templates.

\subsection{The signature of N-variations in the MS}

\begin{figure}
	\includegraphics[width=\columnwidth]{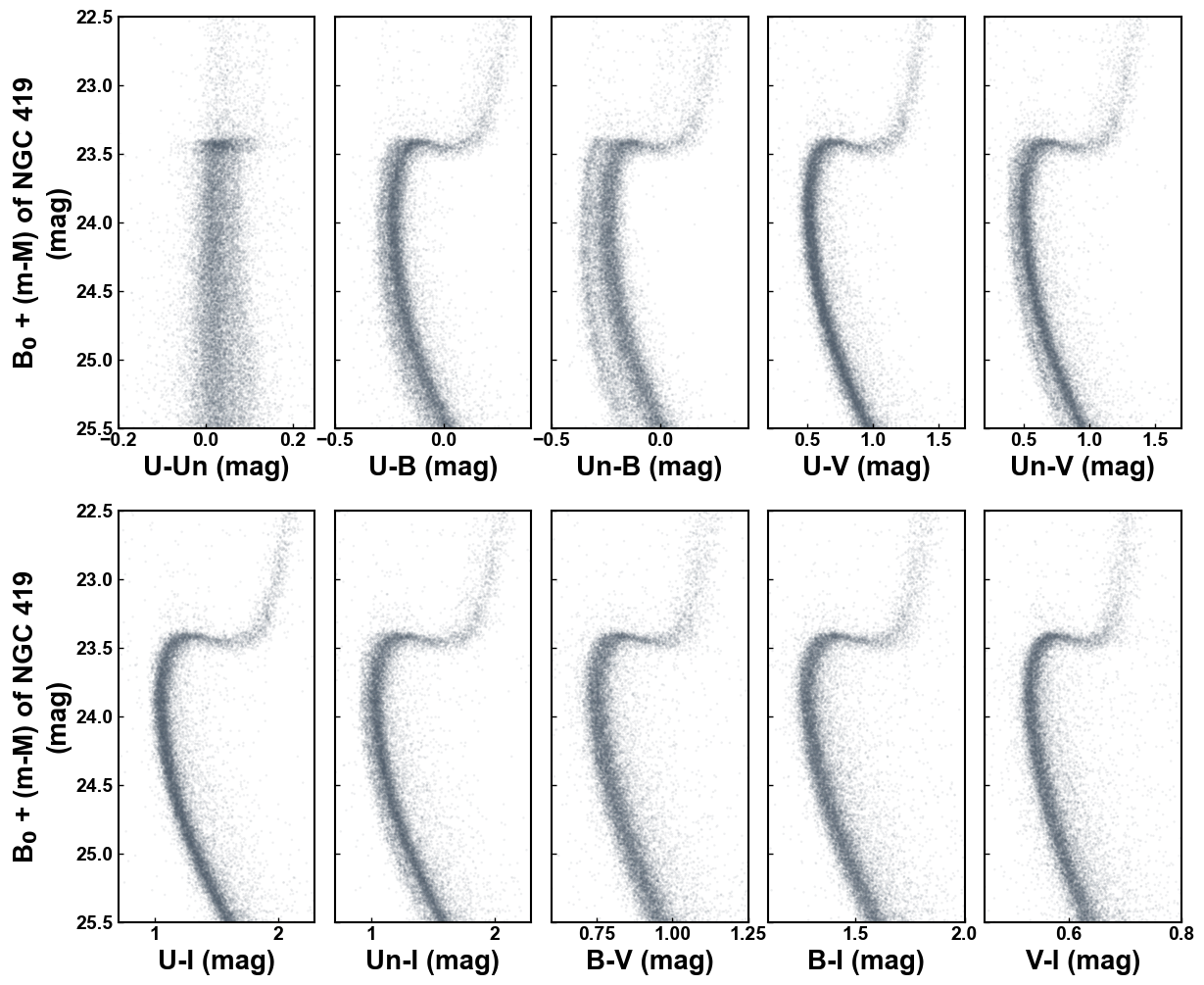}
    \caption{CMDs of 47 Tuc. We plot the CMDs at the same distance as NGC 419 so that stars of the same magnitude will have roughly similar masses. In $Un-B$, $Un-V$ and $Un-I$ colours the difference between the solar scaled (blue MS) and N-rich population (red MS) are very clear.}
    \label{fig:47tuc_ref}
\end{figure}

To start we have chosen 47 Tuc, a Galactic cluster with a similar metallicity to NGC 419 (i.e. [Fe/H]~$\sim-0.7$~dex). In Fig.~\ref{fig:47tuc_ref} we show different CMDs of 47 Tuc in the same bands of our NGC 419 photometry. We applied the distance modulus of NGC 419 to the absolute magnitudes of 47 Tuc in order to simplify the comparison between clusters, i.e. stars at a given apparent magnitude would share roughly similar masses (we adopted $m-M=18.83$~mag from \citealt{Rubele10} for NGC 419 and $m-M=13.266$~mag from \citealt{Gaia18} for 47 Tuc). Note that colours containing the $Un$-band are very effective at separating the solar-scaled and N-enhanced population. These two sub-populations are also appreciable in some colours containing the $U$-band (e.g. $U-B$) however, the distinction between the two is less clear.

\begin{figure*}
	\includegraphics[width=14cm]{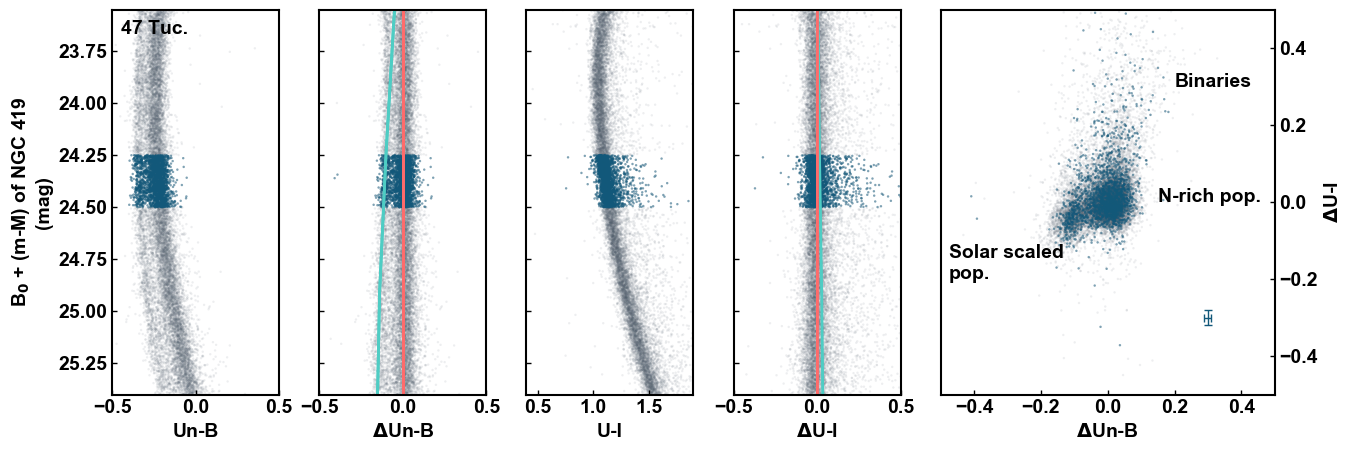}
	\includegraphics[width=14cm]{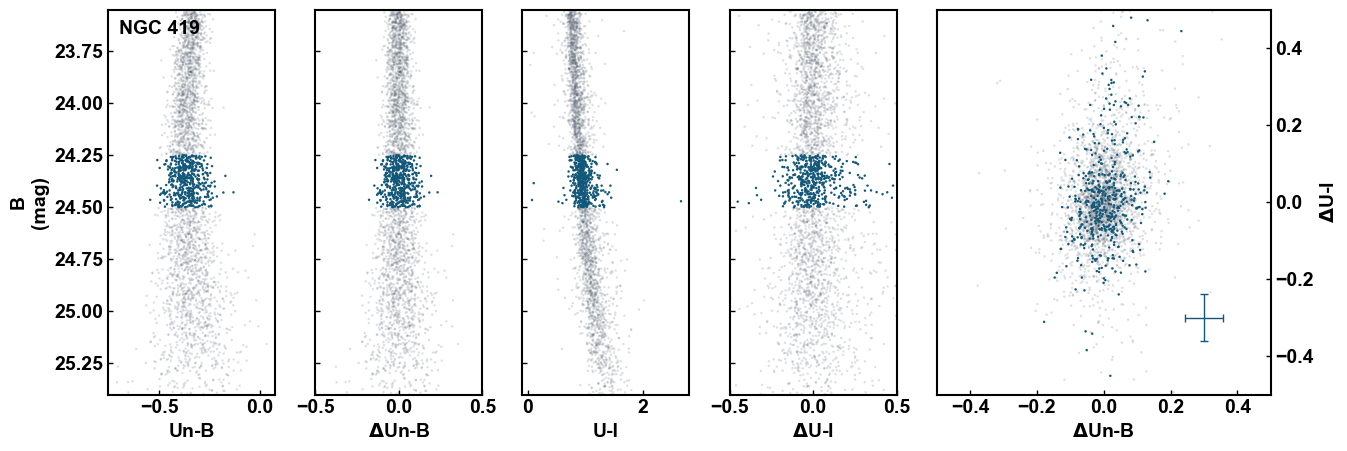}
	\includegraphics[width=14cm]{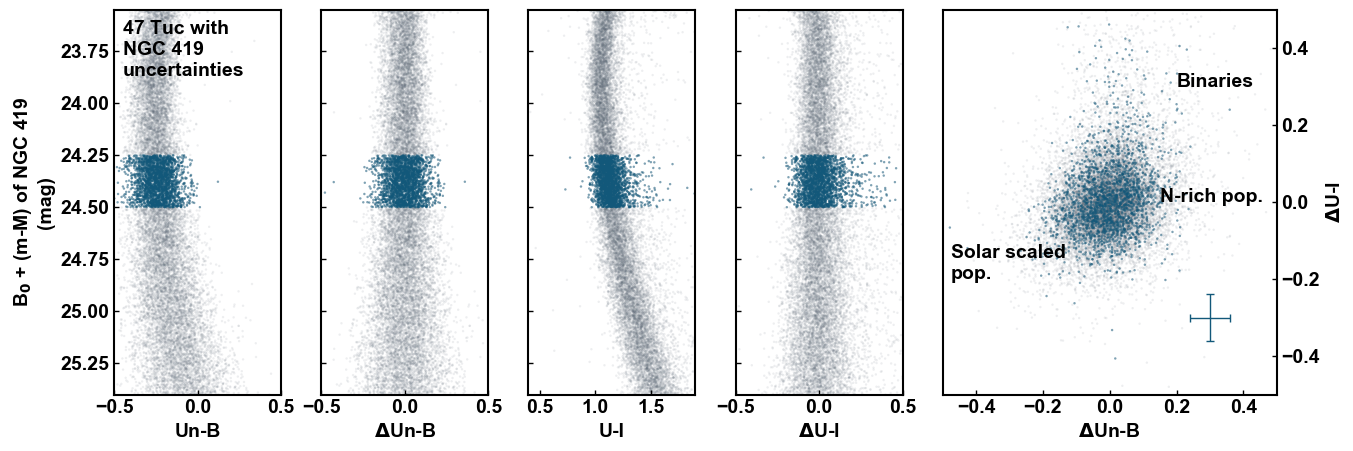}
    \caption{\textbf{Top:} Normalized CMDs and colour-colour plot of 47 Tuc. In this study we focus on stars in the entire range $23.5<B<25.5$ (grey points), however, to facilitate the comparison we highlighted in blue the stars in a narrow magnitude ($24.25<B<24.50$) range in the different panels. The solar scaled (blue) and `extremely' N-rich (red) models are shown for reference in the first row. The solar scaled and N-rich population can be clearly identified in the $Un-B$ CMD, and in the $\Delta (Un-B)$ vs. $\Delta (U-I)$ plot. In $U-I$ the binary stars are very prominent (red tail of the colour distribution), these can also be spotted in the normalized colour-colour plot. \textbf{Middle:} Similar to top panel but for NGC 419. No clear evidence of multiple populations is found by visual inspection of the colour-colour plot. \textbf{Bottom:} Similar to the other panels but now showing a simulation of the CMD of 47 Tuc at the distance of NGC 419. In this case the photometric uncertainties blur the distinctions between the different sub-populations.} 
    \label{fig:colcol}
\end{figure*}

The contrast between sub-populations with different chemical composition can be maximized in a colour-colour plot. For example, in the top panel of Fig.~\ref{fig:colcol} we show the normalized colour-colour plot: $\Delta (Un-B)$ vs. $\Delta (U-I)$ for 47 Tuc stars. These colours were normalized with respect to the mean of the colour distribution at every magnitude, allowing us to compare the relative behaviour of stars in parts of the CMDs with different slopes. The stars from each population occupy distinct loci in $\Delta (Un-B)$ vs. $\Delta (U-I)$ (cf. top right panel of Fig.~\ref{fig:colcol}).

In the middle panels of Fig.~\ref{fig:colcol} we show the equivalent plot for the NGC 419 data. No clear evidence of multiple populations is found by a visual inspection of this colour-colour plot. However, given its distance, the photometric uncertainties of NGC 419 are significantly larger than the ones of 47 Tuc, so it is not clear if one would have been able to pick multiple populations by visual inspection of its colour-colour plot. We can explore this by simulating how the CMD of 47 Tuc would look like if it had the same photometric uncertainties as NGC 419. For this, we added a random Gaussian scatter to our 47 Tuc catalogue in order to match the uncertainties inferred from our AST on the NGC 419 images. The resulting CMDs are shown in the bottom panel of Fig.~\ref{fig:colcol}. Although the stars in the colour-colour plot show a somewhat larger scatter than the expected from the photometric uncertainties, the clear distinction between sub-populations has vanished and now the solar scaled and N-rich population are blended with each other.

\begin{figure}
\begin{center}
	\includegraphics[width=\columnwidth]{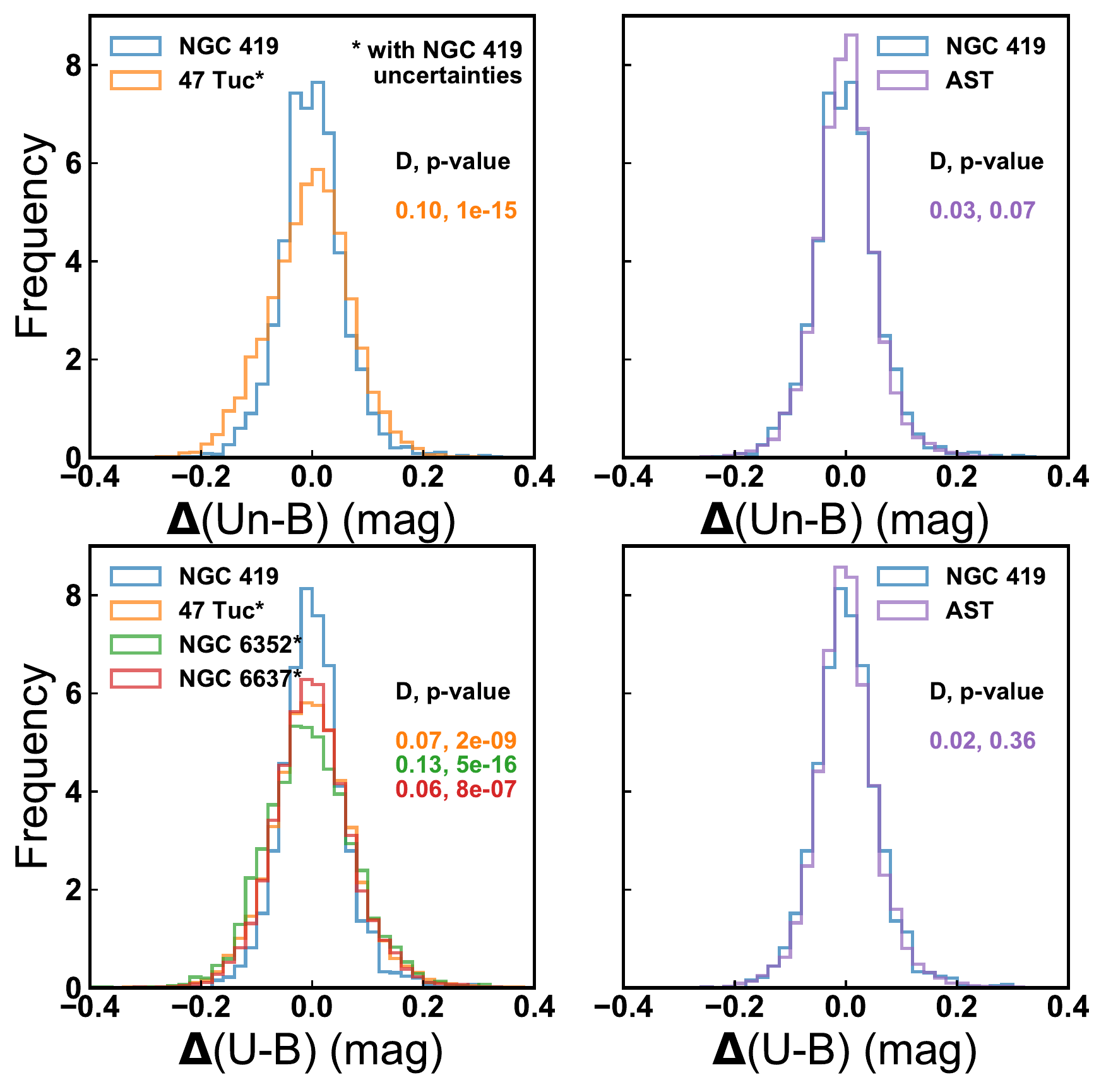}
    \caption{Comparison of the $\Delta (U-B)$ colour distributions of three clusters of similar metallicity as NGC 419. We are focusing on stars in the apparent $B$-magnitude range $23.5-25.5$ (equivalent to $\sim1-0.75$M$_{\odot}$). We report the KS-statistic, D, and p-value for the different cases (see text). As in for the bottom panel of Fig. \ref{fig:colcol} random noise was added to the photometry of 47 Tuc, NGC 6352 and NGC 6637 to match the uncertainties of the NGC 419 data set, see text.}
    \label{fig:ubhist}
\end{center}
\end{figure}

By visual inspection the scatter in $\Delta (Un-B)$ from NGC 419 seems smaller than the one from 47 Tuc, see also the histogram in the top left panel of Fig. \ref{fig:ubhist} and Table \ref{tab}. So if the former were to host multiple populations, the N-variations among its stars would be more subtle than the ones present in 47 Tuc \citep[which show changes up to $\sim1.5-1.7$ dex e.g.][]{Briley04}. One can argue that this would not be unexpected, as mentioned in \S \ref{sec:intro} there is very strong evidence for a correlation between the cluster present day mass and the degree of abundance variations, so in principle this would be entirely consistent with the fact that NGC 419 is about an order of magnitude less massive than 47 Tuc.

That said, we carried out a similar experiment this time comparing the normalised $\Delta (U-B)$ colour distribution of NGC 419 with that of two other galactic globulars (known to host abundance variations) with similar present day mass and metallicities: NGC 6352 and NGC 6637 ($\sim10^5$ M$_{\odot}$~and [Fe/H] $\sim-0.6$ cf. \citealt{Harris96,Baumgardt18}). As mentioned in \S~ \ref{sec:dat}, $Un$ photometry was not available for these targets, however, from Fig. \ref{fig:47tuc_ref} we see that $U-B$ is also a good diagnostic for MPs.

Like for 47 Tuc before, we added random scatter to the catalogues of these clusters to match the uncertainties of NGC 419 data set. We then selected stars with $\sim1-0.75$ M$_{\odot}$~and in the bottom left panel of Fig. \ref{fig:ubhist} we show their $\Delta (U-B)$ colour distribution\footnote{We adopted $m-M=14.43$ and $15.28$ for NGC 6352 and NGC 6637 respectively \citep{Harris96}. Since both clusters are severely affected by extinction, we corrected for differential reddening using the procedure described in \cite{Niederhofer16}.}. The $\Delta (U-B)$ colour distributions of NGC 6352 and NGC 6637 are very similar to the one of 47 Tuc, and significantly wider than the distribution of NGC 419 (cf. Table \ref{tab}).

\begin{table}
\caption{Standard deviation of colour distributions (in mag) of Fig. \ref{fig:ubhist}. We bootstrapped $10^4$ times by resampling 500 stars in each iteration to derive the mean standard deviation of the colour distributions (values in table). The standard error estimated was $\pm0.003$ mag for all distributions except for AST's $\Delta(Un-B)$ which was $\pm0.002$ mag.}
\begin{center}
\begin{tabular}{cccccc}
\hline
$\sigma$ & AST & NGC 419 & 47 Tuc & NGC 6352 & NGC 6637 \\
\hline
$\Delta(Un-B)$ &  0.055 & 0.059 & 0.085 &-- & --\\ 
$\Delta(U-B)$ &  0.057 & 0.057 & 0.076 & 0.079 & 0.073 \\ 
\hline
\end{tabular}
\end{center}
\label{tab}
\end{table}%

That said, there is scaling relation between a globular cluster's mass and the severeness of their abundance variations (cf. \S \ref{sec:intro}). So, the narrow width of NGC 419's colour distributions with respect to 47 Tuc's could be accounted (at least in part) by this relation. Moreover, even though NGC 6352 and NGC 6637 currently have a similar mass to NGC 419, their initial mass might have been significantly larger in the past, which could explain their relatively wide colour distributions\footnote{Both NGC 6352 and NGC 6637 orbit closer to the galactic centre than 47 Tuc  and both NGC 6352 and NGC 6637 have shallow stellar mass functions \citep{Baumgardt19}. Both of these argue towards NGC 6352 and NGC 6637 having lost more stellar mass than 47 Tuc.}. 

For reference, in the right panels of Fig.~\ref{fig:ubhist}, we show a comparison between the $\Delta (U-B)$ and $\Delta (Un-B)$ colours expected from a single stellar population. For this we have used the catalogues generated in the AST. We used these catalogues to create the CMDs, which were normalised in the same way as the real data from the different clusters. The colour distributions from NGC 419, are clearly more consistent with what is expected from a single stellar population than from what is observed in clusters with known abundance variations of similar metallicity.

In each panel of Fig.~\ref{fig:ubhist} we also report the KS-statistic, D, and $p$-value for the different colour distributions\footnote{D is a measure of the maximum distance between the cumulative distributions of NGC 419 and the other clusters. While the $p$-value is the probability of obtaining test results at least as extreme as the results actually observed during the test, assuming that the null hypothesis is correct.}. We can reject the pairwise null hypothesis that the colour distributions of 47 Tuc, NGC 6352 and NGC 6637 and NGC 419 are drawn from the same parent population at the $p=10^{-3}$ level. However, we cannot say the same for the colour distributions of the AST catalogue (i.e. we cannot reject they are `drawn from the same distribution').

\section{Summary and Conclusions}

A growing number of studies have reported that the amplitude of the star-to-star abundance variations characteristic of globular clusters systematically change as a function of global parameters like cluster's mass and metallicity. New spectroscopic and photometric evidence suggest the age of massive clusters also affects the way this phenomenon manifests, pointing towards more subtle signals at younger ages (cf. \S \ref{sec:intro}). 

These studies were focused on RGB stars, however, where evolutionary effects like the first-dredge-up limits any conclusion regarding their \emph{initial} N-composition. This motivates the study of MS stars, i.e. stars which have not undergone these evolutionary effects. The best targets are the same type of low mass stars ($\lesssim 1$ M$_{\odot}$~ -- i.e. stars not affected by fast rotation as this could also affect their surface abundances) that populate the MS of old globulars.

For this work we have analysed different CMDs of low mass stars ($\sim 0.75-1.05$ M$_{\odot}$) of the young massive cluster NGC 419 ($\sim1.4$~Gyr, $\sim2\times10^5$~M$_{\odot}$) to look for evidence for primordial star-to-star N-variations. All things being equal, in colours like $U-B$ and $Un-B$, the presence of such abundance variations would produce broader colour distributions than the ones expected for a cluster with homogeneous abundances.

We use the colour distribution of old clusters with known star-to-star N-variations as empirical templates for our comparison with NGC 419. We find that the main sequence stars of NGC 419 display significantly narrower $U-B$ and $Un-B$ colour distributions than the same stars in our template clusters. Moreover, the colour distributions of NGC 419 seem to be consistent with what is expected from a cluster with homogeneous abundances.\footnote{We note that \cite{Li20} came to similar conclusions with the same dataset in a paper that appeared after the submission of this paper. In summary, they fitted synthetic clusters with different degrees of N-abundance variations to the observations of low mass MS stars of NGC 419. Likewise, they report that their simple stellar population model (i.e. homogeneous abundances) was the best at reproducing the observations. However they cannot exclude small ($\lesssim 0.2$ dex) variations in N abundance.}

These results have very interesting implications. Previous studies of integrated $J$-band spectra of very young ($<30$ Myr) massive ($\sim10^6$ M$_{\odot}$) clusters have found no evidence for the type of abundance variations found in old globulars \citep[cf.][]{cz16,Lardo17}. The integrated near infrared spectra of such young clusters are dominated by the light of red supergiants ($\sim8-35$ M$_{\odot}$), these stars formed and followed a very different evolution than the low mass ($\lesssim1$ M$_{\odot}$) stars found in older globulars. So if the mechanisms responsible for the primordial abundance variations found in old globulars act only in low mass stars, no evidence would be expected in red supergiants.
 A similar argument could be made for stars populating the RGB of young ($\lesssim2$ Gyr) massive clusters, i.e. they are not the same kind of stars as the ones found to host MPs in old globulars (i.e. low-mass stars).

The results presented here suggest that even when comparing `like-with-like' (i.e. the same kind of low-mass/long-lived stars) the abundance variations characteristic of old globulars (e.g. 47 Tuc, NGC 6352 and NGC 6637) appear to be significantly smaller or absent in this young cluster of similar metallicity (NGC 419). That said, we cannot conclude if this effect is driven mostly by its age or its mass. On one hand, there is a well established correlation between the mass of a globular cluster and the strength of the signal of its abundance variations. This could explain in part why the colour distributions of NGC 419 are narrower than those of 47 Tuc ($\sim10^6$ M$_{\odot}$). On the other hand, the study of young $\sim10^6$ M$_{\odot}$~clusters have not produced evidence for MPs as mentioned above, suggesting the age of NGC 419 could also contribute to its apparent chemical homogeneity.

Finally we would like to emphasise that this work represents a case study on a single target. These results should be confirmed independently by different techniques and different targets. If confirmed (i.e. no/subtle-abundance variations), a possible explanation could be that MPs  never occurred in recently formed clusters like NGC 419, because of some special environmental conditions are not satisfied (e.g. low gas pressures/densities in the last couple of Gyr, cf. \citealt{D'ercole16,Elmegreen17}) preventing the mechanisms required for the formation of MPs to operate.

Another alternative is that the process responsible to drive the formation of MPs still operates at young ages, however, it produces smaller fractions of N-enhanced stars and/or the N-enhancements are very subtle. If evidence for (subtle) MPs is eventually found in this cluster, future hypothesis to explain the origin of MPs would need to explain why old clusters ($\sim10$ Gyr) of similar mass and metallicity show stronger signals of abundance variations, i.e. larger fraction of severely N-enhanced stars (e.g. 47 Tuc, NGC 6352 and NGC 6637).

\section*{Acknowledgements}

Support for this work was provided by NASA through grants HST-HF2-51387.001-A (Hubble Fellowship, I.C-Z.) and HST-GO-15062 (V.K-P.) awarded by the Space Telescope Science Institute, which is operated by the Association of Universities for Research in Astronomy, Inc., for NASA, under contract NAS5-26555. J.S.S. is partially supported by funding from the Harvard Data Science Initiative. N.J.B., C.U., S.M. and S.S. gratefully acknowledges financial support from the Royal Society (University Research Fellowship) and the European Research Council (ERC-CoG-646928, Multi-Pop). F.N. acknowledges funding from the European Research Council (ERC) under European Union's Horizon 2020 research and innovation programme (project INTER-CLOUDS, grant agreement no. 682115). C.L. the support from the Swiss National Science Foundation (Ambizione grant PZ00P2\_168065).




\bibliographystyle{mnras}
\bibliography{mnras_template} 




\appendix

\section{Field star subtraction}
\label{sec:decon}

We use a method similar to the one described in \cite{Niederhofer16}. Briefly, one defines a target region (in this case the cluster's centre) and a control field region (outskirts). The idea is that for every star in the control region, we flag a star in the cluster region that has a similar position in the CMD as a likely member of the field -- accounting for the difference in solid angle (area) between the cluster and field regions.

The original method removes the closest star in the target region to the star in the control field. However, here we have implemented a variation where we create a probability distribution function (PDF) for the stars in the target region that is a function of their distance to the control field stars and their respective uncertainties. Then, the star to be flagged as likely member of the field is randomly chosen from this probability distribution.

To obtain the PDF we first need to calculate the distance $d$ in colour magnitude space between a star from the target region and the star from the control region:
$$ d = \sqrt{ (\Delta\mbox{colour})^2 + (\Delta\mbox{magnitude})^2} $$
where $\Delta\mbox{colour}$ and $\Delta\mbox{magnitude}$ are the difference in colour and magnitude between the two stars. We also need the combined uncertainties in colour and magnitude given by:
$${\sigma_c}^2(\mbox{tot}) =  {\sigma_{c}}^2(\mbox{target}) + {\sigma_{c}}^2(\mbox{control}) $$
$${\sigma_m}^2(\mbox{tot}) =  {\sigma_{m}}^2(\mbox{target}) + {\sigma_{m}}^2(\mbox{control}) $$
here $\sigma_c$ and $\sigma_m$ are the respective standard deviations in the colour and magnitude for these stars. With this we calculate the total variance $\sigma^2$ and $\chi^2$:
$$ {\sigma}^{-2} = \frac{1}{{\sigma_c}^2(\mbox{tot})} + \frac{1}{{\sigma_m}^2(\mbox{tot})} $$
$$\chi^2 = d^2/\sigma^2$$
then the PDF for each target star $i$ would be:
$$ \mbox{PDF}_i = \frac{\mathcal{L}_i}{\sum_i \mathcal{L}_i}$$
where,
$$  \mathcal{L}_i = \frac{1}{\sqrt{2\pi\sigma^2_i}}\exp\left(-\frac{\chi^2_i }{2}\right) $$

\begin{figure*}
	\includegraphics[width=14cm]{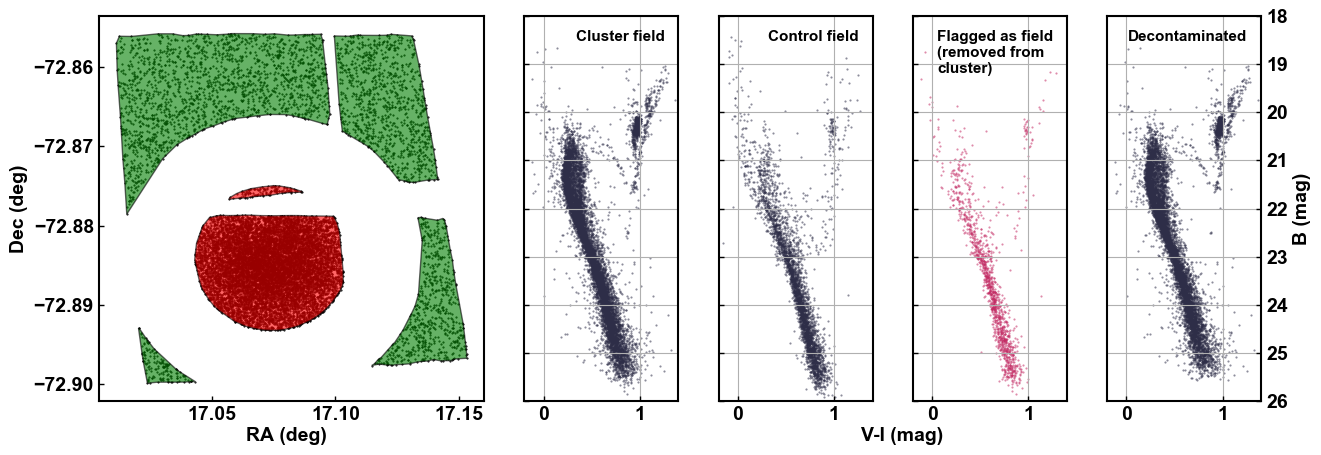}
    \caption{The left panel shows the chosen regions in the sky for the cluster (red) and control field (green). Since the area ratio cluster to control fields is $\sim0.37$, for every star in the control field we flag $\sim0.37$ stars in cluster region as likely field members. The panels to the right show the CMDs of: the centre, outskirts, stars chosen as contaminants (and removed from the cluster catalogue) and the decontaminated CMD. The stars from the latter are the ones used in our analysis (\S~\ref{sec:ms}).}
    \label{fig:cleaning_1ite}
\end{figure*}

In Fig. \ref{fig:cleaning_1ite}~we show the results of this procedure when applied to our NGC 419 data set. In pink we show the stars that have been flagged as likely members of the field and removed from the cluster CMD. The stars in the pink CMD are a representative sample of the stars in the control field CMD. The resulting decontaminated sample (right panel) is the one used in our analysis in \S~ \ref{sec:ms}.

Furthermore, this method has the advantage that it enable us to calculate the probability of a given star to belong to the field. This is achieved by carrying out multiple realisations of this procedure and keeping track of the frequency each star is flagged as a likely field member. This is not possible in the Niederhofer et al. method as that method will always identify the same stars as field members if one repeats the procedure.

\begin{figure*}
	\includegraphics[width=12cm]{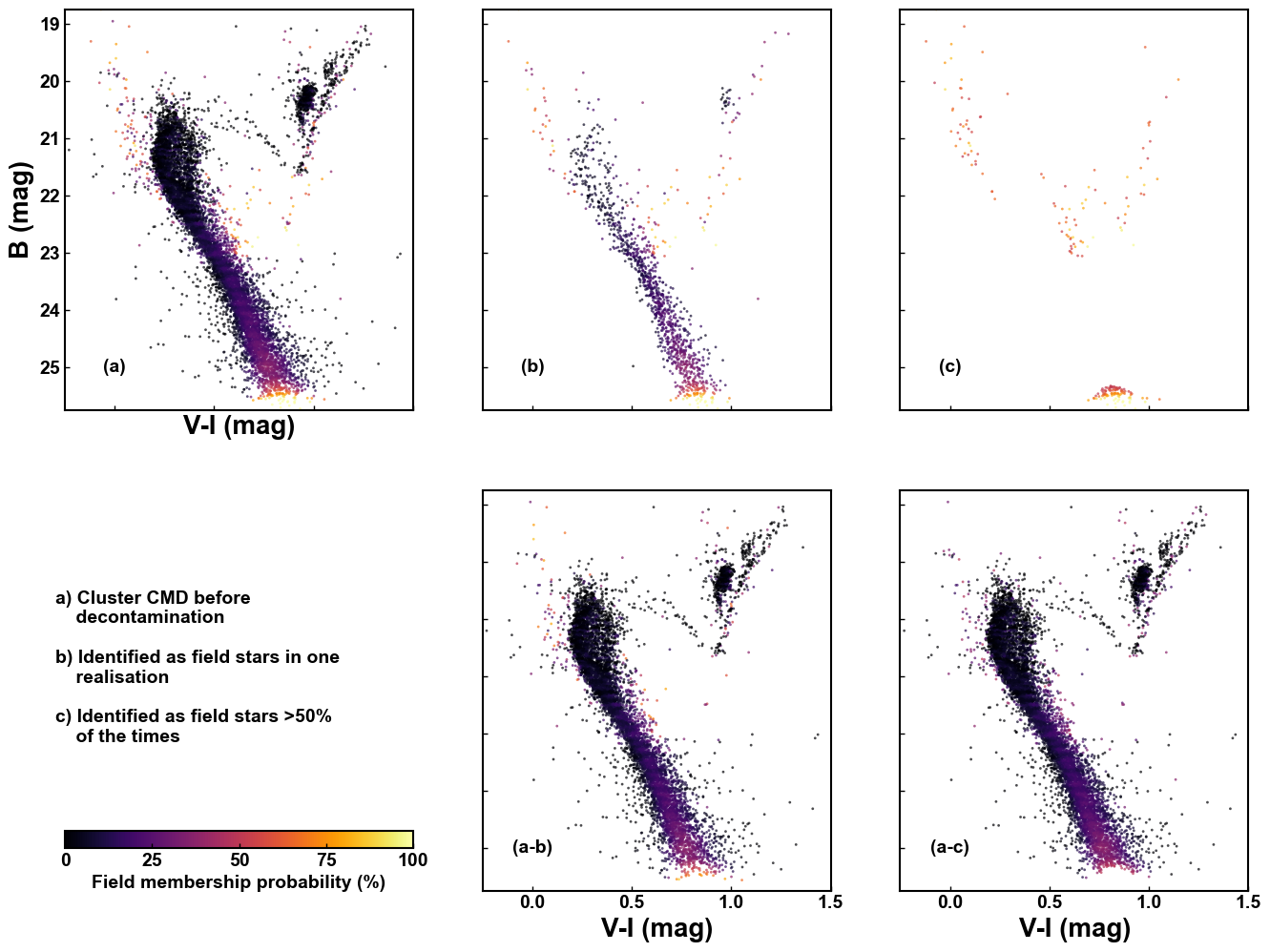}
    \caption{CMDs of the central region of NGC 419 with stars colour coded by their field membership probability derived from 1000 realisations of our decontamination technique. (a) CMD of cluster field. (b) Stars flagged as field in one realisation of our method (i.e. pinks CMD of Fig. \ref{fig:cleaning_1ite}). (c) Stars identified as field members $>50\%$ of our 1000 realisations. In the bottom row, we find the decontaminated CMDs by subtracting from (a) stars from (b) and (c). The panel (a-b) show the same CMDs as the rightmost panels of Fig. \ref{fig:cleaning_1ite}, but for reference, now the stars are colour coded with their field membership probability. See text for discussions.}
    \label{fig:multi}
\end{figure*}

For example, in Fig. \ref{fig:multi} we present different CMDs where the stars are colour coded by their probability  to belong to the field. There probabilities were determined from 1000 realisations of our method and counting how many times a given star was flagged as a field member. In panel (a) we show the cluster region CMD before decontamination. Panels (b) and (c) present two alternative samples of field star contaminants, and below decontaminated CMDs using those samples -- panels (a-b) and (a-c).

Panel (b) shows the same stars shown in pink in Fig. \ref{fig:cleaning_1ite}, these stars were flagged as field stars in one realisation of our method. By removing these stars from the CMD show in panel (a), we will obtain a relatively \emph{pure} sample of cluster stars across all evolutionary phases. However, this method assumes that the stellar populations found in the control region are uniformly distributed across the entire field of view \emph{and}, that there are no cluster stars in the control region. Clearly this last assumption is not satisfied in our NGC 419 data set (cf. control field CMD in Fig.~\ref{fig:cleaning_1ite}), where we still have a non negligible contribution of the cluster in our control region. This means that for a region like the cluster turn-off would tend to `over subtract' stars -- the dark points in the turn-off region of (b) of Fig. \ref{fig:multi} are likely cluster members.

On the other hand, if one were interested for example in a more \emph{complete} sample of stars in the cluster turn-off (i.e. less over-subtraction), that could be achieved by simply setting a field membership probability threshold. In panel (c) of Fig. \ref{fig:multi} we show an example of the stars with $>50\%$ chance of being field members. Removing those stars from the CMD we would minimize the over-subtraction of stars in the turn-off or RGB, however, this comes at the expense of the \emph{purity} of the sample (it would have more contaminants).

We note that although this method folds in the uncertainties the photometry of both cluster and control field, if a certain part of the CMD is not well populated in your control/background sample, it will be very unlikely flagged as field member in your cluster CMD, e.g. outliers at the faint end ($\mbox{B}>23$ mag) of the MS. As one would expect, the better the background sample one feeds in, the better result one gets out.

Having said that, given our limitations from the restricted HST pointings for this target a better alternative for the control field is not available. So any effort to decontaminate this data set from field stars will be compromized by this.

\begin{figure}
\begin{center}
	\includegraphics[width=\columnwidth]{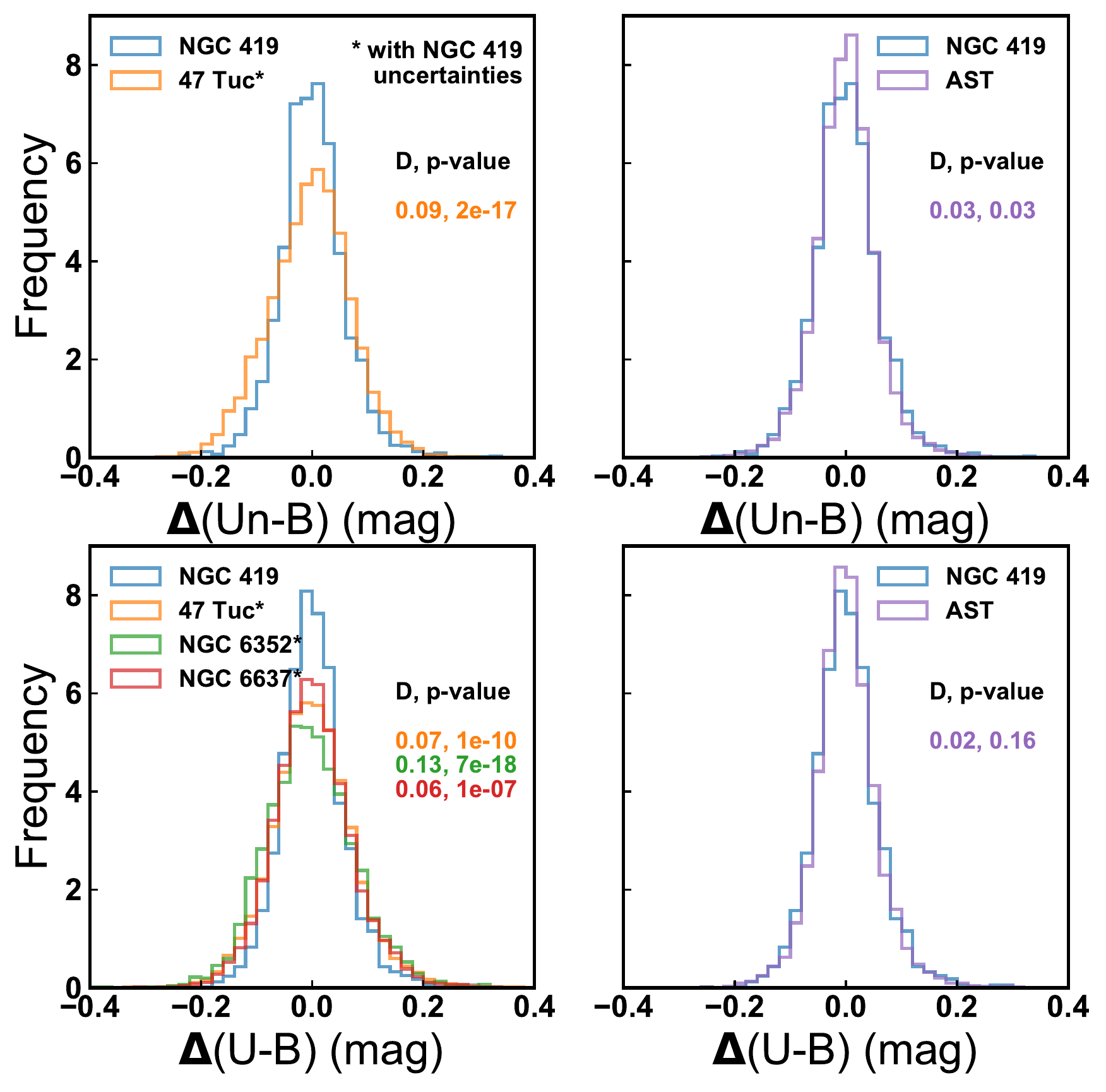}
    \caption{Similar to Fig. \ref{fig:ubhist}. But instead of decontaminating our sample using the procedure outlined in Fig. \ref{fig:cleaning_1ite} (i.e. one realisation of our method), here we removed from the CMD of NGC 419 all stars with $>50\%$ probability to belong to the field -- i.e. stars from panel (a-c) in Fig. \ref{fig:multi}. The results remain the same, i.e. the colour distributions of the other cluster are significantly wider than NGC 419.}
    \label{fig:ubhist50}
\end{center}
\end{figure}

Finally, for our particular science case the choice of field decontamination method did not affect our conclusions, and the same results were found by subtracting stars by setting a $50\%$ cut in field membership probability (cf. Fig. \ref{fig:ubhist50}), or by simply sampling of stars according to the field region CMD like in Fig. \ref{fig:cleaning_1ite} and removing those from the cluster sample.

\bsp	
\label{lastpage}
\end{document}